\documentclass[preprintnumbers,amsmath,12pt,amssymb,floatfix,superscriptaddress,nofootinbib]{article}

\def\myname#1{\begin{center}{\large #1}\end{center}\vspace{-0.13cm}}
\def\myplace#1#2{\small\begin{flushleft}\textit{#1}\\
\texttt{#2}\end{flushleft}}

\def\myclassification#1{\small\noindent
Keywords :  #1\vspace{0.5cm}}
\usepackage{amsmath}
\usepackage[table]{xcolor}
\usepackage{color}
\usepackage{amsfonts}
\usepackage{amssymb}
\usepackage{breqn}
\usepackage{array}
\usepackage{graphicx}
\usepackage[utf8]{inputenc}
\usepackage[numbers]{natbib}
\usepackage{hyperref}
 
 \date{}

\begin{document}
\title{A Brief Thermodynamic Study For Four Dimensional Einstein Gauss Bonnet Black Holes Using Fractalised Barrow Entropy}
\maketitle

\maketitle

\myname{Ritabrata~Biswas*\footnote{biswas.ritabrata@gmail.com~~;~~\text{Orchid}~:~0000-0003-3086-892X} and Satyajit~Pal*,**\footnote{satyajitpal1995@gmail.com~~;~~\text{Orchid}~:~0009-0009-2099-4501}}
\vspace{0.5cm}
\myplace{*Department of Mathematics, The University of Burdwan, Burdwan-713104, India\\ **Department of Mathematics, Dr. Bhupendra Nath Dutta Smriti Mahavidyalaya, Hatgobindapur, Purba Bardhaman-713407, India}{}

\begin{abstract}
Higher dimensional Gauss-Bonnet gravity can be particularized to a four dimensional case either using the Glavan, D. and Lin, C. type \citep{glavan2020einstein} limiting method or by the Hordenski type \citep{gurses2007gauss}
metric compactification procedure. Depending on ADM mass and Gauss Bonnet coupling parameter $\alpha_{GB}$, a black hole solution is prescribed \citep{hennigar2020taking}. The phrase which is responsible for divergence at some finite values of radial coordinate stays inside a square root and thus softens the divergence. A quantum gravity affected surface formula for the black hole is chosen. Fractalized entropy thus formed is known as the Barrow Entropy\citep{ladghami2024barrow}. Global nature for such an entropy formula is discovered. Temperature is calculated with and without different quantum corrections. Stability with such temperature structures are analyzed. To support this, sign changes in specific heat, occurrences of double points in free energy, sign changes and jumps in derivatives of free energy etc are thoroughly discussed. Probable rise of phase transitions are pointed.
\end{abstract}

\myclassification{Einstein Gauss Bonnet black holes, Fractalized entropy, Phase transitions}

\section{Introduction}
A black hole is a scrap of space-time trapped inside a causal horizon popularly called as the ``event horizon". The escape speed of such a dense object turns enumerously high for even an electromagnetic wave to escape \cite{wald1994quantum}. This made it hard to fetch any internal information of the concerned wrapped region. Consequently, an external observer is left only to continue the concerned research with the externally observable properties. If the object under consideration is a static black hole, such parameters are the black hole's mass $M$, charge $Q$ and the angular momentum $J$ \cite{bekenstein2004black,kiefer2006quantum} etc. A set of black hole’s thermodynamic laws does exist to characterize the whole mechanics of black holes.

The Zeroth law states about the uniform distribution of the surface gravity $\kappa$ throughout a black hole's event horizon \cite{wald1994quantum,bardeen1973four}. If $ \delta M $ amount of change in a black hole's mass takes place which is responsible for  $ \delta A_g $, $ \delta J $, $ \delta Q $ changes in surface, angular momentum and charge respectively, said changes are related as
\begin{equation}\label{1st_law_0th_Equn}
\kappa \delta A_g = 8 \pi G_N \left(\delta M - \Omega \delta J - \phi \delta Q \right)~~~~,
\end{equation}
where $ \Omega $ and $ \phi $ respectively are the angular velocity and electrostatic potential of the black hole. $ G_N $ is Newton's gravitational constant. Equation (\ref{1st_law_0th_Equn}) is the  basement  of first law of black hole physics.

The surface area of a black hole’s event horizon $(A_g = 4\pi R_g^2)$ is an increasing function of time. This quantity due to its ever increasing nature, can be treated as an analogous entropy. Even when two black holes (say with mass $M_1$ and $M_2$) merge, the resultant entropy (say $A_{g, res}$) will be larger than the algebraic sum of individual entropies (say $A_{g,1}$ and $A_{g,2}$) \cite{kiefer2006quantum,hawking1971gravitational,hawking1972black,christodoulou1970reversible}. 
\begin{dmath}\label{result_entropy}
    A_{g, res} = 4\pi R_{g, res}^2 = 4\pi \left\{ 2 G_N \frac{(M_1 + M_2)}{c^2} \right\}^2 = 4\pi \left( \frac{2G_N M_1}{c^2}\right)^2 + 4\pi \left( \frac{2G_N M_2}{c^2}\right)^2 + 4\pi \left( \frac{2 G_N}{c^2} \right)^2 2M_1 M_2  > A_{g, 1} +A_{g ,2}~~~.
\end{dmath}
Second law of black hole thermodynamics has its origin here. Finally, for a stationary black hole, the surface gravity $ \kappa $ can not vanish by applying a finite number of processes \cite{israel1986third}. This supports the third law of thermodynamics.

 A fractal horizon is imagined in the article \cite{barrow2020area} by attaching some (say set $S_1$ of)  small spheres to touch the outer surface of a Schwarzschild black hole of mass  $M$ and radius $ R_g = \frac{2 G_N M}{c^2} $, $c$ being the speed of light. Second set $S_2$ of smaller spheres touching each of  these small sphere’s surface (from $S_1$ set) are considered and so on. Following Koch Snowflake boundary \cite{barrow2020area} method, hierarchically smaller touching spheres are constructed and their surfaces will now build up a new boundary. Let us assume that each step of smaller scale intricacy laws to the inclusion of $N$ spheres with $ \lambda $  times smaller radius than the previous spheres' to which the new set is attached tangentially. Hence the recursion of radii is  $ r_{n + 1} = \lambda r_{n} $  at $n^{th}$ step, $ r_0 = R_g $ . 
 
If we allow up the smaller spheres to touch the surface, the total volume $ V_{\infty} $ of the summed-up black hole after an infinite number of steps, turns to be
\begin{equation}\label{fractal_volume}
 V_{\infty} = \sum_{n = 0}^{\infty} N^n \frac{4 \pi}{3} \left(\lambda ^n R_g\right)^3 = \frac{4 \pi R_g ^3}{3} \sum_{n=0}^{\infty} \left(N \lambda ^3\right)^n \rightarrow \frac{4 \pi R_g ^3}{3 \left(1 - N \lambda ^3 \right)} > \frac{4 \pi R_g ^3}{3}~~~~.
\end{equation}
as $N \rightarrow \infty$ (while $N \lambda^3 <1$). After finding the finiteness of the volume $ V_{\infty}$ , we also check infinite steps' total surface area as 
\begin{equation}\label{limiting_fractal_Area}
A_{g,\infty} = \sum_{n = 0} ^{\infty} N^n 4 \pi \left(\lambda ^n R_g \right)^2 = 4 \pi R_g ^2 \sum_{n =0}^{\infty} \left(N \lambda ^2 \right)^n > 4 \pi R_g ^2~~~.
\end{equation}
However, if $ N \lambda ^2 < 1 $, the total surface area diverges to
\begin{equation}\label{non_limiting_fractal_volume}
 A_{g,\infty} = \frac{4 \pi R_g ^2}{1 - N \lambda ^2} ~~~~.
\end{equation}
But to keep the volume finite with an infinite area, $ N \lambda ^2 < 1 $ should be chosen and we reach to a constraint on $N$ as
\begin{equation}\label{N_constraints}
 \frac{1}{\lambda ^2} < N < \frac{1}{\lambda ^ 3}~~ .
\end{equation}
At limit $ N \to \infty $, the area of the extended surface diverges, entropy turns infinitely large and turns meaningless as a physical indicator. On the other hand, convergence to a finite limit too has an entropy greater than the spherical Schwarzschild surface area.

This can be realized more physically when we think about the number of spheres to fit around the bigger sphere of the last iteration. If we take a $2D$ slice passing through the center of the bigger surface, a circle with a radius $ R_g + r $ will be formed to pass through the centers of the smaller ones. If the maximum number of smaller circles is $N$, then $N$ times the smaller diameters $2r$ will be equal to the perimeter $ 2 \pi (R_g + r) $ of the larger circle, i.e, $ N^r = \pi (R_g + r) $ . 

Hence we pass a constrain of the radii multiplier $\lambda $ (as $ r = \lambda R_g $ ) by $ N \leq \pi \left(\lambda^{-1} + 1 \right) $ . Although the true bond  will be a $3D$ one, this $2D$ slice estimatation is indicative and concordant with the constrain $ \frac{1}{\lambda ^2} < N < \frac{1}{\lambda ^ 3} $ .

For a Schwarzschild black hole, its surface area $ A_g = 4 \pi R_g ^2 $ determines its entropy as 
\begin{equation}\label{Entropy}
 S = \frac{A_g c^3}{4 G_N \hbar} \approx \frac{A_g}{A_{pl}}~~~~ ,
\end{equation}
$ A_{pl} $ being Planck area. Hence entropy is simply the number of Planck areas accumulated inside the horizon area. Now the increased value is actually derived from the area theorem, $ \frac{dA_g}{dt} \geq 0 $, with increasing complexity and determination required to emphasize the horizon structure.

As the luminosity is proportional to $ A_g T_H ^4 $, where $ T_H \propto \frac{1}{M} $ is the black hole temperature and if the increment in area took place by an $ \alpha $ multiplicator $ ( \alpha \geq 1) $, then the increased area will go through rapid evaporation and the Hawking lifetime $ t_{bh} \propto \frac{M^3}{\gamma ^2} $ will fall. If no upper bound of $ \gamma $ is provided, a primordial black hole will explode rapidly without leaving any remnant for today's observation.

If the surface of the extended black hole is pure fractal, the surface area will vary with radius as $ R_g^{2 + \Delta},~ 0 \leq \Delta \leq 1 $. For the limiting cases\\
$ \Delta = 0 $ :   simplest horizon structure
\\
$ \Delta = 1 $ :   one geometric dimension raised area.
\\
Here the corresponding entropy is coined as the Barrow entropy which will vary as 
\begin{equation}\label{Barrow_Entropy}
S_B \approx \left(\frac{A_g}{A_{pl}} \right) \approx \left(\frac{A_g}{A_{pl}} \right)^{\frac{2 + \Delta}{2}} .
\end{equation}
To physically interpret, for our observable universe inside the particle horizon, we take for present time, $ A_g \approx \left(c t_0 \right)^2 $ and present cosmic age to $ \approx 10^{17}s $ and obtain 
\begin{equation}\label{Barrow_Entropy_Physical Example}
S_{univ} \approx \left(\frac{10^{17}}{10^{-43}} \right)^{2 + \Delta} \approx 10^{120 \left(1 + \frac{\Delta}{2} \right)}  \approx 
	\begin{cases}
    		  10^{120} & \text{, for smooth spacetime structure}\\
     	  10^{180} & \text{, for most fractralised structure}
	\end{cases}~~~~~~.
\end{equation}
In the view of this, the quest for achieving a more complete theory of quantum gravity becomes increasingly convincing and provides additional perspectives to further deepen the understanding of black holes. Indeed, use of the fractal structure model to study black holes helps to understand the relation between quantum gravity, space-time and gravity. A number of studies have talked about Barrow’s entropy on black holes and cosmology. For the sake of example, we mention regarding the article  \cite{ladghami2024barrow} for Barrow’s entropy in cosmology and the articles\cite{petridis2023barrow, ladghami2k24barrow} for the traditional and holographic thermodynamics of black holes with Barrow's fractal structure.

It is true that the fractal structure is connected to well-known theories of quantum gravity. As described in the Barrow model, fractals can illustrate space-time foam and exists beyond conventional quantum gravity models. In particular, a space-time foam fractal is predicted by the loop quantum gravity model together with the famous result of a quantum space-time micro-structure \cite{rovelli2008loop, carlip2023spacetime}. Also, in string theory, one has interpreted the quantum production of $D$-branes as a type of space-time foam \cite{hartnoll2006spacetime, ellis1997quantum}. Moreover, the pioneering work of Barrow provided an explanation for the physically admissible incorporation of a fractal structure into black hole thermodynamics. This foam is produced by oscillations at Planck length scales, where space-time is not smooth, but foamy. Thus, the event horizon area of black holes is also characterized by a fractal structure, which means that the temperature and entropy of black holes have to be modified.

The modified Friedmann equations in the context of the Barrow entropy turns \cite{phong2022baryogenesis}
\begin{equation}
H^{2-\Delta}-\Delta H^{-\Delta}H'\frac{1}{a}=\frac{8\pi G_N}{3}\sum_i \bar{\rho}_i ~~\text{and}
\end{equation}
\begin{equation}
H^{-\Delta}\left[\left(\Delta-2\right)H'\frac{1}{a}-3H^2+\Delta\left\{\frac{H''}{H}-(1+\delta)\left(\frac{H'}{H}\right)^2\right\}\frac{1}{a^2}\right]-\Delta H^{-\Delta}H'\frac{1}{a}=\frac{8\pi G_N}{3}\sum_i \bar{p}_i ~~~~,
\end{equation}
where $H=\frac{\dot{a}}{a}$ is the Hubble parameter, $a\equiv a(t)$ is the scale factor,  $\bar{\rho}_i$ and $p_i$ respectively represent the energy density and pressure of the $i$-th component of the universe. In the Barrow-Tsalli's cosmological framework, the entropy is a combination of Barrow and Tsalli's entropies, leading to a modified primordial gravitational wave(PGW) spectrum. The relic density of PGWs in this context is given by \cite{barrow2020area}
\begin{equation}\label{GW}
\Omega_{GW}(\tau, \kappa)=\Omega_{GW}^{GR}(\tau, \kappa)\left[\frac{a_{hc}}{a_{hc}^{GR}}\right]\left[\frac{H_{hc}}{H_{hc}^{GR}}\right]^2~~~~,
\end{equation}
where $\Omega_{GW}^{GR}(\tau, \kappa)$ is the primordial gravitational wave relic density in standard GR and the ratios involving $a_{hc}$ and $H_{hc}$ account for the differences introduced by the Barrow-Tsalli's modifications\cite{jizba2024imprints}.

Using (\ref{GW}) and the modified PGW spectrum in Barrow-Tsalli's cosmology, different signatures are followed in the studies of advanced GW detectors.
Big Bang Observer (BBO), found to act sensitive to low frequency PGWs, is able to detect suppressed spectrum for $\Delta>0$ and also to constrain $\Delta$ upto an approximate order of $\mathcal{O}\left(10^{-3}\right)$ \cite{jizba2024imprints}. Laser Interferometer Space Antenna(LISA) is sensitive to intermediate frequencies which enables it to detect PGWs with enhanced spectra for $\Delta<0$, providing constraints on $\Delta$ in the negative range \cite{jizba2024imprints}. Square Kilometer Array(SKA) is capable in timing array observations and detects PGWs in such frequency ranges where enhancements due to negative $\Delta$ are significant \cite{jizba2024imprints}. Lastly, Pulsar Timing Arrays (PTAs) have been used to place constraints on $\Delta$, excluding values as low as $\Delta\lesssim-5\times 10^{-2}$ based on recent data \cite{jizba2024imprints}.

Depending on the black hole temperature formula, we write
\begin{equation}\label{temperature1}
T_H = \frac{\hbar c^3}{8 \pi k_B G_N M} = \frac{\kappa \hbar}{2 \pi k_B c}~~~~ ,
\end{equation}
where $ \kappa = \frac{c^4}{4 M G_N} $ for Schwarzschild black hole, Using the idea of specific heat $ C_V = \left(\frac{\partial E}{\partial T_H} \right)_{V,P} $, the expression turns 
\begin{equation}\label{specific heat}
C_V = - \frac{\hbar c^5}{8 \pi G_N K_B T_H^2} ~~~~.
\end{equation}
This implies black holes get cooler when energy is added and get heated up with the extraction of energy. Though second law $ dA_g \geq 0 $ prevents that, black holes' evaporation and Hawking radiation supports this issue.

More recently,
it has been noticed that the conventional second law of thermodynamics is missing a ``work done" term. Some propose \cite{ladghami2024barrow} the cosmological constant $ \Lambda $ (an imprint of AdS gravity) as the gravitational analog of pressure as it determines the background curvature of the concerned space-time. Along with one conventional prefecture  $ -\frac{1}{8 \pi G_N} $, we write $ P \equiv \frac{\Lambda}{8 \pi G_N} $.

Stephen Hawking and Don Page \cite{hawking1983thermodynamics} studied the thermodynamics of Anti-deSitter (AdS) black holes to notice a transformation between pure thermal AdS space and Schwarzschild AdS black hole.

If a black hole thermodynamic volume is defined as $ V = \left(\frac{\partial M}{\partial P} \right)_{S, Q, J} $, then it is followed that phase transition of charged AdS black hole remarkably coincides with the Van der Wall's liquid gas phase transition \cite{belhaj2013thermodynamical,li2014effects,ladghami2024black,cai2013pv,gunasekaran2012extended}. 

In AdS space,  different black hole solutions are observed to pass through phase, between small and large black holes \cite{kubizvnak2012p} , stretched quintessence \cite{ladghami2024thermodynamics}, multiple critical points \cite{tavakoli2022multi},  polymer-type phase transitions \cite{dolan2014isolated} and Joule Thompson expansion \cite{okcu2017thermal} etc. Holographic thermodynamics, a work on the framework of AdS-CFT correspondence \cite{maldacena1999large}, deals with thermal studies in CFT \cite{cong2022holographic}.

In this article, we are motivated to study the effects of fractalization of entropy onto a 4D EGB gravity black hole's thermodynamics. Mathematical origin and structure of such a black hole softens the divergence. This leads to a classical resolution of the singularities. Barrow entropy is important because it provides a phenomenological framework to incorporate quantum gravity induced geometric deformations into black hole and cosmological thermodynamics. It offers a way to study the implications of a fractal or nonsmooth space-time structure on fundamental gravitational and cosmological phenomena. Fate of such a model in thermodynamic perspective is to be studied in this article.

In the next section, a brief introduction to a $4D$ Einstein Gauss Bonnet black hole(4DEGB hereafter) will be given. Thermodynamic study of a 4DEGB gravity will be done in the third section. Finally, a brief discussion and conclusion to this article will be given in the last section.

\section{4D Einstein Gauss Bonnet Gravity Black Holes}
We recall $D$ dimensional Einstein Gauss Bonnet action given by 
\begin{equation}
{\cal S}^{(D)}_{EGB}=\frac{1}{16\pi G_N^{(D)}c^4}\int d^Dx\sqrt{-g} \left[R-2\Lambda+\alpha_{GB} {\cal G}\right]~~~~,
\end{equation}
where $\alpha_{GB}$ is Gauss Bonnet regulatory parameter and the Gauss-Bonnet term
$${\cal G}=R^2-4R_{\mu\nu} R^{\mu \nu}+R_{\mu\nu\rho\sigma} R^{\mu \nu\rho\sigma}~~~~$$
is followed to contribute to the equation of motion only when $D>4$. $R$, $R_{\mu\nu}$ and $R_{\alpha \beta \gamma \delta}$ are Ricci scalar, Ricci tensor and Riemann tensor respectively. If we opt four dimensions, it becomes a total derivative which hardly has any impact on the local dynamics.

Glavan, D. and Lin, C.\cite{glavan2020einstein} have rescaled the Gauss Bonnet coupling as $\alpha_{GB} \rightarrow \frac{\tilde{\alpha}}{D-4}$ and redefined the action as
\begin{equation}
{\cal S}^{(D)}_{EGB}=\frac{1}{16\pi G_N^{(D)}c^4}\int d^Dx\sqrt{-g} \left[R-2\Lambda+\frac{\tilde{\alpha}}{D-4} {\cal G}\right]~~~~.
\end{equation}
${\cal G}$ is topological in $4D$. However, the prefactor diverges as $D\rightarrow 4$. So we will take the variation before applying the limit as 
\begin{equation}\label{variation}
\delta{\cal S}^{(D)}_{EGB}=\frac{1}{16\pi G_N^{(D)}c^4}\int d^Dx\sqrt{-g} \left[\delta R-2\Lambda+\frac{\tilde{\alpha}}{D-4} \delta {\cal G}\right]~~~~.
\end{equation}
$ \delta {\cal G}$ contributes to a finite term to field equation even if when $D=4$. From \ref{variation}, we find 
\begin{equation}
G_{\mu\nu}+\Lambda g_{\mu\nu}+\tilde{\alpha} H_{\mu\nu}=8\pi T_{\mu\nu}~~~~,
\end{equation}
where $G_{\mu\nu}=R_{\mu\nu}-\frac{1}{2}Rg_{\mu\nu}$ is the Einstein's tensor and $H_{\mu\nu}$ is the effective correction coming from the Gauss Bonnet term in the limit $D\rightarrow 4$ given as
\begin{equation}
H_{\mu\nu}=2 R R_{\mu\nu}-4R_{\mu\alpha}R^{\alpha}_{\nu}-4R_{\mu\alpha\nu\beta}R^{\alpha\beta}+2R_{\mu}^{\alpha\beta\gamma}R_{\gamma\alpha\beta \nu}-\frac{1}{2}g_{\mu\nu}{\cal G}~~~~,
\end{equation}
In $4D$ GR, this tensor identically vanishes. But in $4D$ EGB theory, this comes as a finite residual effect of $D\rightarrow 4$ limit. Theoretical consistency of this approach is debated though.

Another alternative(Horndeski-type) to dissolve the glitch is popular as Kaluza-Klein like reduction which uses conformally rescaled matrices\cite{gurses2007gauss}. Compactification of the metric causes dimensional reduction of a higher dimensional Lovelock gravity and leads to effective $4D$ theory is reached. Here Gauss Bonnet term couples to a scalar field and in $4D$, the action reads as
\begin{equation}\label{EGB_action}
S_{EGB} = \frac{1}{2\kappa} \int d^4 x \sqrt{-g} \left( R - 2 \Lambda  +  \alpha_{GB} \left[ \phi \mathcal{G}\,  + 4 G_{\mu \nu}\nabla ^{\mu} \phi \nabla^{\nu} \phi  - 4 \left(\nabla \phi \right)^2 \square\phi + 2 \bigl\{ \left(\nabla \phi \right)^2 \bigl\}^2 \right] \right) + {\cal S}_m~~~~ ,
\end{equation}
where $ \kappa = 8 \pi G_N c^{-4},~ \phi $ is the dimensionless scalar field, $ \alpha_{GB} $ represents the 4DEGB coupling constant (in length squared units), $ {\cal S}_m $ indicates the matter action. The equation \ref{EGB_action} remains invariant under a transformation where the scalar field \cite{clifton2012modified} is shifted by a constant value $ \mathcal{C} $.
\begin{equation}\label{invariance_term}
\phi \rightarrow \phi + \mathcal{C}~~~~.
\end{equation}
For the scalar field, the field equation is provided by \cite{hennigar2020taking}
\begin{dmath}\label{4DEGB_Field_Equation}
\mathcal{G}\, - 8 G_{\mu\nu}\nabla^{\mu}\nabla^{\nu}\phi  -  8R_{\mu\nu}\nabla^{\mu}\phi\nabla^{\nu}\phi + 8(\square\phi)^2 - 8\nabla_{\mu}\nabla_{\nu}\phi\nabla^{\mu}\nabla^{\mu}\phi - 16\nabla_{\mu}\nabla_{\nu}\phi\nabla^{\nu}\phi\nabla^{\mu}\phi - 8(\nabla\phi)^2 \square\phi = 0 ,
\end{dmath}
whereas the following field equations result from changing the action in relation to the metric
\begin{dmath}\label{4DEGB_Field_Equation2}
G_{\mu\nu} + \Lambda g_{\mu\nu} + \alpha_{GB} \left[ \phi H_{\mu\nu} - 2 R \left\{ \nabla_{\mu} \nabla_{\nu} \phi  + \left( \nabla_{\mu} \phi \right) \left( \nabla_{\nu} \phi \right) \right\} + 8 R^{\rho}_{(\mu}\nabla_{\nu)} \nabla_{\rho}\phi + 8 R^{\rho}_{(\mu}\nabla_{\nu)}\phi\nabla_{\rho}\phi - 2 G_{\mu\nu} \left\{ \left( \nabla\phi \right)^2 + 2\square\phi \right\} - 4 \left\{ \nabla_{\mu}\nabla_{\nu}\phi + \left( \nabla_{\mu}\phi \right) \left( \nabla_{\nu}\phi \right) \right\} \square \phi - \left\{ g_{\mu\nu} \left( \nabla\phi \right)^2 - 4 \left( \nabla_{\mu} \phi \right) \left( \nabla_{\nu}\phi \right) \right\} \left( \nabla\phi \right)^2 + 8 \nabla_{\rho} \nabla_{(\mu} \phi \left( \nabla_{\nu)}\phi \right) \nabla^{\rho} \phi - 4 g_{\mu \nu} R^{\rho \sigma} \left\{ \nabla_{\sigma} \nabla_{\rho} \phi + \left( \nabla_{\sigma} \phi \right) \left( \nabla_{\rho} \phi \right) \right\} + 2 g_{\mu \nu} \left( \square \phi \right)^2 - 2 g_{\mu \nu} \left( \nabla_{\sigma} \nabla_{\rho} \phi \right) \left( \nabla^{\rho} \nabla^{\sigma} \phi \right) - 4 g_{\mu \nu} \left( \nabla_{\rho} \nabla_{\sigma} \phi \right)\left( \nabla^{\rho} \phi \right) \left( \nabla^{\sigma}\phi \right) + 4 \left( \nabla_{\mu} \nabla_{\rho} \phi \right) \left( \nabla_{\nu}\nabla^{\rho}\phi \right) + R_{\mu \rho \nu \sigma} \left\{ \nabla^{\rho} \nabla^{\sigma} \phi + \left( \nabla^{\rho} \phi \right) \left( \nabla^{\sigma}\phi \right)\right\} \right] = \frac{8\pi G_N}{c^4}T_{\mu\nu} 
\end{dmath}
provided the stress energy tensor takes the form
\begin{equation}\label{Stress_Energy}
T_{\mu\nu} := -\frac{2}{\sqrt{-g}}\frac{\delta {\cal S}_m}{\delta g^{\mu\nu}} ~~.
\end{equation}
$\square = \nabla_{\mu}\nabla^{\mu}$ is the d'Alembert operator .

It disappears in four dimensions or less in an identical manner. The cosmological constant that we shall use in the following is zero.

4D EGB theory features a precise vacuum solution, characterized by a line element expressed as
\begin{equation}\label{metric}
ds^2 = -f(r)(cdt)^2 + \frac{dr^2}{f(r)} + r^2 (d\theta^2 + \sin^2 \theta d\varphi^2)~~~~,
\end{equation}
in which the lapse function and the derivative of the scalar field $ \phi $ are respectively provided by \cite{hennigar2020taking}
\begin{equation}\label{solved metric}
f(r) = 1 + \frac{r^2}{2\alpha_{GB}}\left(1 - \sqrt{1+\frac{8\alpha_{GB} G_N M}{c^2 r^3}}\right)~~~~\text{and}~~~\frac{d\phi}{dr} = \frac{\sqrt{f} - 1}{r\sqrt{f}}~~~~,
\end{equation}
$M$ being an integrating constant. Since this solution is asymptotically flat, $M$ can be considered as the mass of a black hole that does not rotate. By solving the equation $ f(r_h) = 0 $, where $ r_h $ represents the radius of the event horizon, we may get the expression for the mass of the black hole.
\begin{equation}\label{mass}
M = \frac{c^2 \alpha_{GB}}{2 G_N r} + \frac{c^2 r}{2 G_N}~~~.
\end{equation}
In the next section, we will study the thermodynamics built up with 4DEGB gravity black hole for a fractalized entropy.
\section{Thermodynamics of $4DEGB$ Gravity Black Holes for Fractalized Entropy}
Now finding the event horizon and Cauchy horizon from equation (\ref{mass}), 
\begin{equation}\label{horizons}
r_{h/C} = M\pm \sqrt{M^2-\frac{c^4}{G_N^2}\alpha_{GB}^2}~~~.
\end{equation}
When $r_h$ and $r_C$ are the event horizon and the Cauchy horizon, respectively. Set up which leads to two horizons is coined as a double horizon structure. Classically, Reissner-Nordstrom and Kerr black holes possess such horizons. Among these horizons, the outer event horizon does not let light to escape. Another singular boundary, namely the inner Cauchy horizon exists where due to blue shifted infalling radiation, predictability of space-time breaks down. This leads to mass inflation under perturbation\cite{poisson1990internal}.

If we check near horizon symmetries and holography, such double horizons often give rise to enhanced near horizon conformal symmetries which are crucial in AdS/CFT and Kerr/CFT correspondence. To understand black hole microstates and entropy from a quantum gravity perspective, this symmetry plays important role \cite{guica2009kerr}. Quantum back reaction indicates how quantum fields affect the space time geometry itself. Stress energy from quantum fluctuations becomes enormous near the inner horizons which turns into strong reaction which destabilizes the inner horizon. Sometimes the horizon is removed\cite{poisson1990internal}. In certain String theory compactification, external black holes with double horizons are used to match microscopic and macroscopic entropy relations \cite{bekenstein2008bekenstein}.

We find that the product $ \frac{c^4}{G_N^2}\alpha_{GB}^2$ of these two horizons is free of the mass of the corresponding black hole. We can express the entropy as follows, accounting for the deformations caused by the black hole’s region due to quantum gravity
\begin{equation}\label{fractal_entropy}
S_{B_{h/C}} = (\pi r^2)^{1+ \frac{\delta}{2}}~~~~.
\end{equation}
We follow that the entropy product $S_{B,h}S_{B,C}=\left(\pi\alpha_{GB}^4 \frac{c^8}{G_N^4}\right)^{1+\frac{\delta}{2}}$ is free of the mass term. Hence, fractal entropy of 4DEGB gravity is a global property. Hence the entropy product law is a universal law for concerned kind of black hole and the fractal entropy\cite{ansorg2009inner, cvetivc2011universal}. 

More degrees of freedom is added due to this additional fractal microstructure on the surface. Dark energy models in cosmological contexts are constructed using holographic modification inspired by the Barrow fractalized entropy \cite{saridakis2020barrow}. If we think of a $2D$ smooth ($\delta=0$) holographic film representing the universe's boundary, the information is evenly encoded. On the other hand, for a crinkled or fractalized ($\delta> 0$), the film incorporates more surface area per unit length which allows more information to be encoded. Simply, this is the interpretation of Barrow correction \cite{saridakis2020barrow}.

From the mass equation \eqref{mass} and the entropy expression \eqref{fractal_entropy}, writing $S_{B_{h}}=S_B$ for simplicity, we obtain
\begin{equation}\label{temperature}
T_H = \left( \frac{\partial M}{\partial S_B} \right) = \frac{c^2}{2 \sqrt{\pi} (2 + \delta) G_N S_B} \left(S_B ^{\frac{1}{2 + \delta}} - \alpha_{GB} \pi S_B^{- \frac{1}{2 + \delta}}\right) ~~~~.
\end{equation}
\begin{figure}[h!]
    \centering
     ~~~~~~~~~~~~~~~~~~~Fig ~1a ~~~~~~~~~~~~~~~~~~~~~~~~~~~~~~~~~~~~~~~~~~~~~~~~~~~~Fig~1b~~\\
    \includegraphics[width=0.45\linewidth]{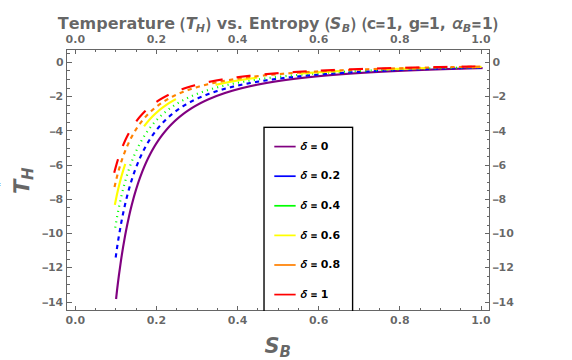}~\includegraphics[width=0.45\linewidth]{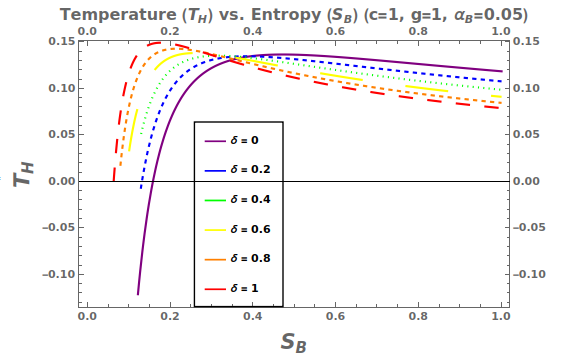}\\  
    ~~~~~~~~~~Fig~1c~~~~~~\\
    \includegraphics[width=0.45\linewidth]{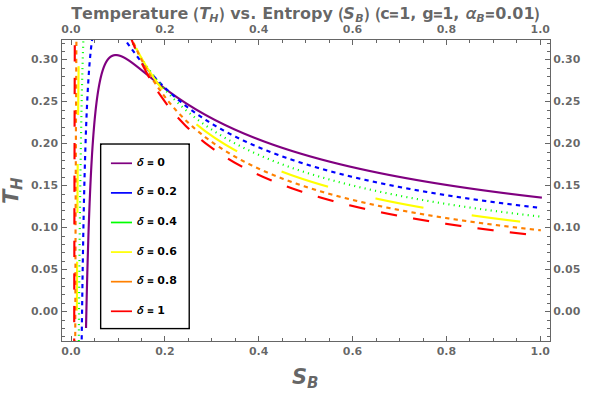}\\
    Figure Caption : Fig 1a , 1b and 1c are Temperature vs. entropy plots for different values of Barrow's index, $\delta$ ($0\leq\delta\leq 1$) in 4DEGB gravity at Gauss Bonnet coupling parameter $\alpha_{GB}=1, ~0.05~ \& ~0.01$ respectively.
    \label{fig:EGB_temperature}
\end{figure}
In Figure 1a to 1c, Hawking temperature versus entropy is plotted for different values of $\delta$ ($0 \leq \delta \leq 1$). The Gauss-Bonnet parameter $\alpha_{GB}$ is kept equal to 1, 0.05, 0.01 in Fig. 1a, 1b, and 1c respectively.

When the Gauss Bonnet parameter $\alpha_{GB} = 1$, i.e, the effect of Gauss Bonnet term is high, temperature of simple black hole structure $(\delta =0)$ is ever increasing. First, with a rapid increment which until a point  $S_B = S_{Bc}(\delta)$. For $S_B > S_{Bc}(\delta)$, the rate of increment falls. As we increase $\delta$, i.e, fractal entropy is considered, temperature is found to increase fast to reach a value at $S_B = S_{Bc}(\delta)$ $[ S_{Bc}(\delta_1) < S_{Bc}(\delta_2)~,~ \delta_2 > \delta_1]$ and stays almost constant valued after that. As we reduce $\alpha_{GB}$'s value to 0.05 in Fig 1b, a local maxima is followed to take place at some $S_B = S_{Bc}(\delta)$. For $S_B < S_{Bc}$, temperature increases with a rapid speed and for  $S_B > S_{Bc}$, temperature reduces with comparatively slower speed. For this case,  $ S_{Bc}(\delta_1) < S_{Bc}(\delta_2)$ also when $\delta_2 <\delta_1$. The pattern says that when $\alpha_{GB}$ is reduced to 0.01 in Fig 1c for $\delta = 0$, temperature is first increasing to reach a local maxima and then falls slowly. But as $\delta$ is increased, these two prominent parts get disjoint. The first increasing part represents an unphysical one and only the decreasing branch is of interest here. 

For $ \alpha_{GB} \leq 0 $, two horizons can be obtained in addition to for $ M > \frac{c^2 \sqrt{\alpha_{GB}}}{G_N} = M_{min} $ if $ \alpha_{GB} > 0 .$ The outer event horizon \cite{charmousis2022astrophysical} is found at
\begin{equation}\label{Event horizon}
R_h = \frac{G_N M}{c^2} + \sqrt{\frac{G_N^2 M^2}{c^4} - \alpha_{GB}}~~~.
\end{equation}
This is less than its Schwarzschild counterpart for  $ \alpha_{GB} > 0 $. This theory has other branches of symmetric solutions, but it is the only one that is flat at infinity and does not have any naked singularities \cite{fernandes2021black}. As a result, the line element (\ref{metric}) with metric function (\ref{solved metric}) will show the spacetime outside a neutron star that is round.

Expressing $ f(r) = 1 + \frac{2 \phi (r)}{c^2} $, gravitational force can be obtained per unit mass in 4DEGB caused by a spherical body
\begin{equation}\label{f}
\vec{f} =- \frac{d\varphi}{dr} \hat{r} = - \frac{c^2 r}{2\alpha_{GB}} \left(1 - \frac{c^2 r^3 + 2 \alpha_{GB} G_N M}{c^2 r^3 + 8 \alpha_{GB} G_N M} \sqrt{1+ \frac{8 \alpha_{GB} G_N M}{c^2 r^3}} \right) \hat{r} ~~~,
\end{equation}
whose magnitude is less than that of its Newtonian $ \left(\alpha_{GB} = 0 \right) $ equivalent $ \left(\vec{f_N} = \frac{- G_N M \hat{r}}{r^2} \right) $ for $ \alpha_{GB} > 0 $. Equation (\ref{f}) disappears for $ r = ( \frac{\alpha_{GB} G_N M}{c^2})^{\frac{1}{3}} $, Nevertheless, this happens at an $r$ value below the outer horizon of the related black hole (as noted in (\ref{Event horizon})). So as long as $ \alpha_{GB} > 0 $, the gravitational pull outside any spherical object stays attractive even if it is not as strong as in GR.

 When $ \alpha_{GB} < 0 $, the gravitational force is more appealing than it is in GR. The empirical constraint is obtained, nevertheless, from the necessity that atomic nuclei not be protected by a horizon \cite{charmousis2022astrophysical}. The constraint
\begin{equation}\label{alpha}
\alpha_{GB} \geq -10^{-30} m^2 ~~~,
\end{equation}
causes the gravitational effects to be completely imperceptible. For pragmatic reasons, negative $ \alpha_{GB} $ can be left out of our analysis. 

Using LAGEOS satellites, an upper limit for the coupling constant 
\begin{equation}\label{alpha2}
0 < \alpha_{GB} \leq 10^{10} m^2
\end{equation}
has been obtained \cite{fernandes20224d}. Preliminary calculations based on recent gravitational wave(GW) statistics indicate that these limitations may be much more stringent \cite{fernandes20224d} which is given by
\begin{equation}\label{alpha3}
 0 < \alpha_{GB} \leq 10^{7} m^2 .
\end{equation}
However, a proper computation still needs to be done. 

We recheck the expression (\ref{f}) for a large $r$  as
\begin{equation}
\sqrt{1+\frac{8\alpha_{GB} G_N M}{c^2r^3}}=1+\frac{4\alpha_{GB} G_N M}{c^2r^3}-\frac{8\alpha_{GB}^2G_N^2M^2}{c^4r^6}+\dots .
\end{equation}
Simplifying
\begin{multline}
\frac{c^2r^3+2\alpha_{GB} G_N M}{c^2r^3+8\alpha_{GB} G_N M}\left(1+\frac{4\alpha_{GB} G_N M}{c^2r^3}\right)^{\frac{1}{2}}\approx \left(1-\frac{6\alpha_{GB} G_N M}{c^2r^3}\right)\left(1+\frac{4\alpha_{GB} G_N M}{c^2r^3}\right)\\
\approx \left(1-\frac{2\alpha_{GB} G_N M}{c^2r^3}\right)+\mathcal{O}\left(r^{-6}\right) ~\text{as}~ r\rightarrow \infty    
\end{multline}
Plugging into the force equation (\ref{f}),
\begin{equation}
\vec{f}(r)\approx -\frac{c^2r}{2\alpha_{GB}} \left[1-\left(1-\frac{2\alpha_{GB} G_N M}{c^2r^3}\right)\right]\hat{r}=-\frac{c^2r}{2\alpha_{GB}}\frac{2\alpha_{GB} G_N M}{c^2r^3}\hat{r}~~~~.
\end{equation}
This shows leading order Newtonian gravity is reconstructed and the next order correction would be 
$$\delta(r)=\frac{\text{correction~ terms}}{\frac{G_NM}{r^2}}=\mathcal{O}\left(r^{-3}\right)~~~~.$$
On the other hand, scalar tensor gravity predicts $\vec{f}(r)=-\{1+\delta(r)\}\frac{G_N M}{r^2}\hat{r}$.
If we compare with scalar tensor theory, especially the Brans-Dicke like models, the gravitational coupling is modified to
$$G_{eff}(r)=G_N\left(1+\delta(r)\right)~~~~\delta(r)\sim \beta^2exp\{-m_{\phi}r\}~~~~,$$
for a scalar mass $m_\phi$ and coupling  strength $\beta$, the corresponding force term
$$\vec{f}(r)=-(1+\delta(r))\frac{G_NM}{r^2}\hat{r}$$
and the correction term $\delta(r)\sim \frac{1}{r^n}$ for massless scalar fields.

Depending on the scalar coupling and dynamics, $n$ is observed to be $1,~2$ and $3$. So our force law mimics scalar tensor gravity in the weak field limit and implies a scalar degree of freedom with nontrivial back reaction, just like in dirty black hole scenario(Bacharia-Bronnikov-Melkinov-Bekenstein)\cite{alford1998qcd,
sotiriou2012black, anabalon2012asymptotically}

We will now extract and compare the structure of the force correction term $\delta(r)$ that arises due to scalar field effects in both cases.

Here, we can follow up a quantum modification known as Generalized Uncertainty Principle which, motivated by theories like String theory or loop quantum gravity, near the Planck scale, modifies Heisenberg uncertainty principle as 
\begin{equation}
\delta x\delta p\geq \frac{\hbar}{2}\left[1+\beta \left(\delta p\right)^2\right]~~~~,
\end{equation}
here $\beta=\beta_0\frac{l_p^2}{\hbar^2}$. The Planck length $l_p=\sqrt{\frac{\hbar G_N}{c^3}}$ is used along with a dimensionless parameter $\beta_0$ (typically assumed to be $\sim ~1$). This leads to a temperature correction \cite{gangopadhyay2014generalized}
\begin{equation}
T_{H,GUP}\approx T_H\left[1+\frac{\beta_0}{\pi^2}\left(\frac{M_P}{M}\right)^2\right]\approx \frac{M_P^2c^2}{8\pi M k_B}\left[1+\frac{\beta_0^2M_P^2}{\pi^2M^2}\right]~~~~,
\end{equation}
where $M_P=\frac{\hbar c}{G_N}$ is Planck mass. 
\begin{figure}[h!]
    \centering
     ~~~~~~~~~~~~~~~~~~~Fig ~1d ~~~~~~~~~~~~~~~~~~~~~~~~~~~~~~~~~~~~~~~~~~~~~~~~~~~~Fig~1e~~\\
    \includegraphics[width=0.45\linewidth]{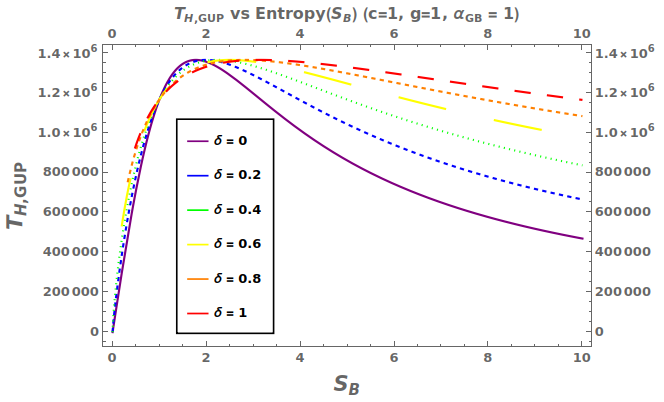}~\includegraphics[width=0.45\linewidth]{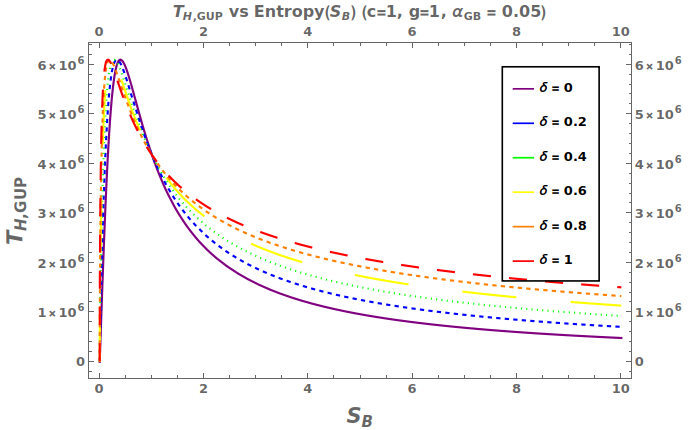}\\  
    ~~~~~~~~~~Fig~1f~~~~~~\\
    \includegraphics[width=0.45\linewidth]{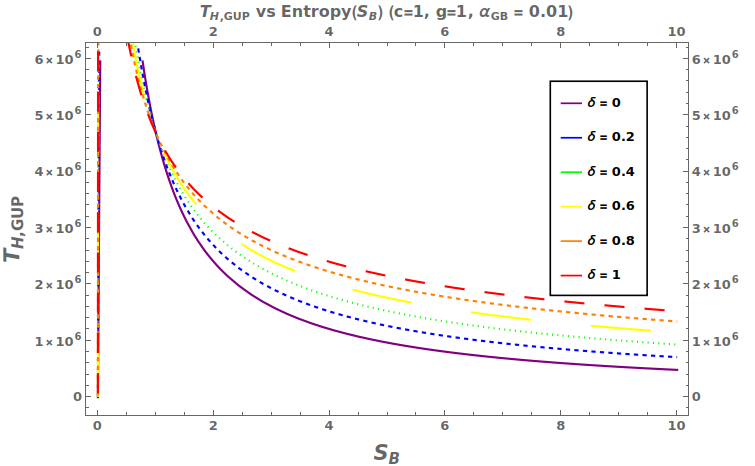}\\
    Figure Caption : Fig 1d , 1e and 1f are $T_{H,GUP}$ vs. entropy plots for different values of $\delta$ ($0\leq\delta\leq 1$) in 4DEGB gravity at Gauss Bonnet coupling parameter $\alpha_{GB}=1, ~0.05~ \& ~0.01$ respectively.
    \label{fig:T_{H,GUP}}
\end{figure}
$T_{H,GUP}$ is plotted vs $S_B$ in figures 1d $(\alpha_{GB}=1)$, 1e$(\alpha_{GB}=0.05)$ and 1f$(\alpha_{GB}=0.01)$.In 1d, as quantum effect is included, temperature increases with almost a constant slope for $\delta=0$. As Barrow's index grows larger, GUP corected temperature increases with a high slope first and then the rate slows down. Comparatively this is increasing for large domain of $S_B$ than $T_H$.

As $\alpha_{GB}$ turns 0.05, temperature falls quickly for larger $S_B$. Quantum effect opposes that of Barrow's index as for different $\delta$ even, temperature merge with each other. When $\alpha_{GB}$ is very low (=0.01) GUP corrected temperature does not differ much than the non corrected case.

Another quantum correction to black hole temperature arises from quantum loop corrections. This idea incorporates virtual particle effects and back reaction on the space-time geometry divided in two primary branches : matter loop corrections and graviton loop corrections\cite{frolov1996one}.

Here the corrected temperature takes the form 
\begin{equation}
T_{H, Q}=T_H\left(1+\alpha_Q \frac{\hbar }{M^2}+\dots\right)~~~~,
\end{equation}
the constant $\alpha_Q$ depends on the particle content, spin and regularization scheme. 

\begin{figure}[h!]
    \centering
     ~~~~~~~~~~~~~~~~~~~Fig ~1g ~~~~~~~~~~~~~~~~~~~~~~~~~~~~~~~~~~~~~~~~~~~~~~~~~~~~Fig~1h~~\\
    \includegraphics[width=0.45\linewidth]{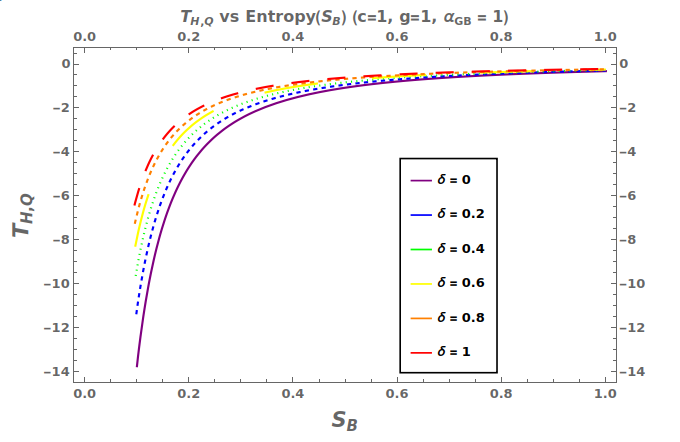}~\includegraphics[width=0.45\linewidth]{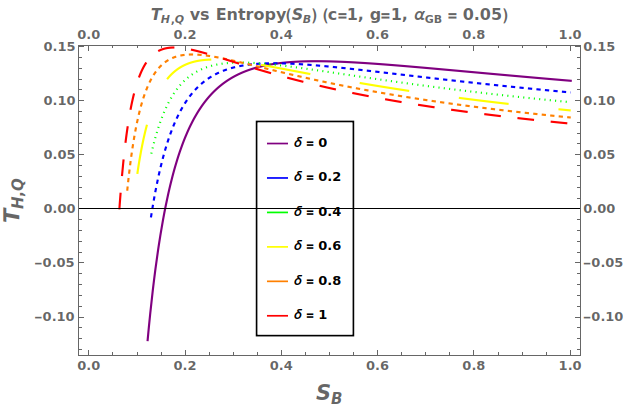}\\  
    ~~~~~~~~~~Fig~1i~~~~~~\\
    \includegraphics[width=0.45\linewidth]{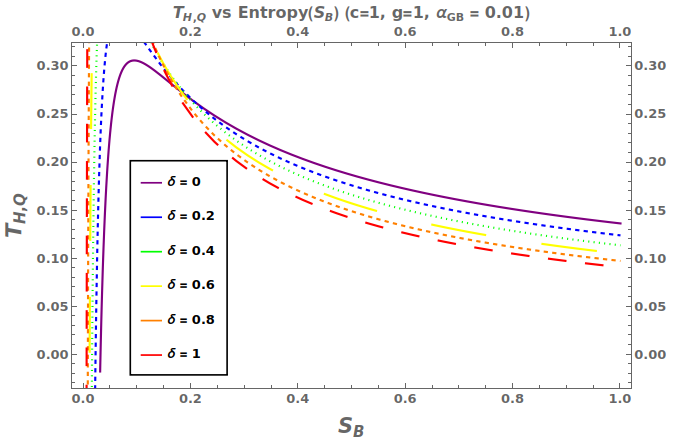}\\
    Figure Caption : Fig 1d , 1e and 1f are $T_{H,Q}$ vs. entropy plots for different values of $\delta$ ($0\leq\delta\leq 1$) in 4DEGB gravity at Gauss Bonnet coupling parameter $\alpha_{GB}=1, ~0.05~ \& ~0.01$ respectively.
    \label{fig:T_{H,Q}}
\end{figure}
Quantum corrected temperature is plotted in 1g-i respectively for $\alpha_{GB}=1,~0.05,~0.01$. Basic nature of the plots does not differ much than that of simple 4DEGB case.

However, these corrections are merely noticeable for small black holes. This section analyzes how the fractal structure affects thermodynamic processes, different types of phase transitions and black hole stability. In order to investigate black hole stability and evaluate the influence of the fractal structure, we compute the heat capacity using the subsequent formula,
\begin{equation}\label{speecific_heat2}
C = \frac{\left(2 + \delta \right) S_B \left(S_B^{\frac{2}{2 + \delta}} - \pi \alpha_{GB} \right)}{ \left( 3 + \delta \right) \pi \alpha_{GB}- S_B^{\frac{2}{2 + \delta}} \left(1 + \delta \right)} ~~~.
\end{equation}
\begin{figure}[h!]
    \centering
       ~~~~~~~~~~~~~~~~~~~Fig ~2a ~~~~~~~~~~~~~~~~~~~~~~~~~~~~~~~~~~~~~~~~~~~~~~~~~~~~Fig~2b~~\\
    \includegraphics[width=0.45\linewidth]{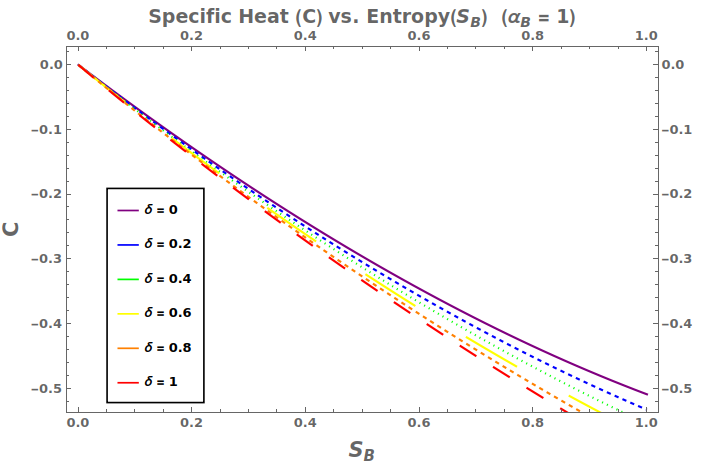}~\includegraphics[width=0.45\linewidth]{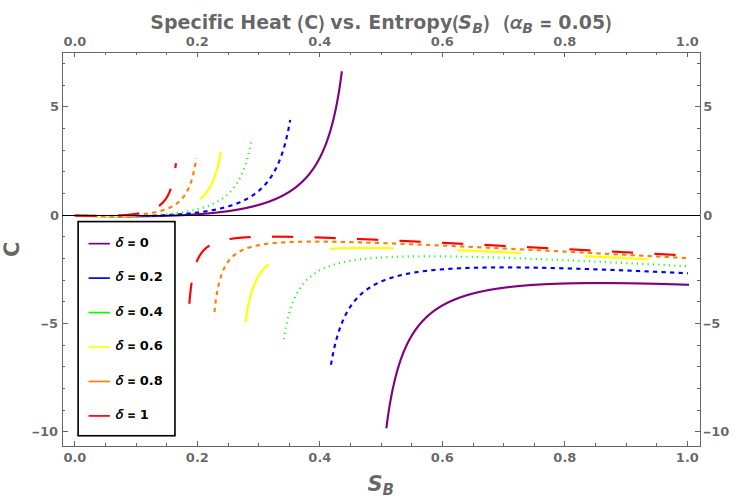}\\  
    ~~~~~~~~~~Fig~2c~~~~~~\\
    \includegraphics[width=0.45\linewidth]{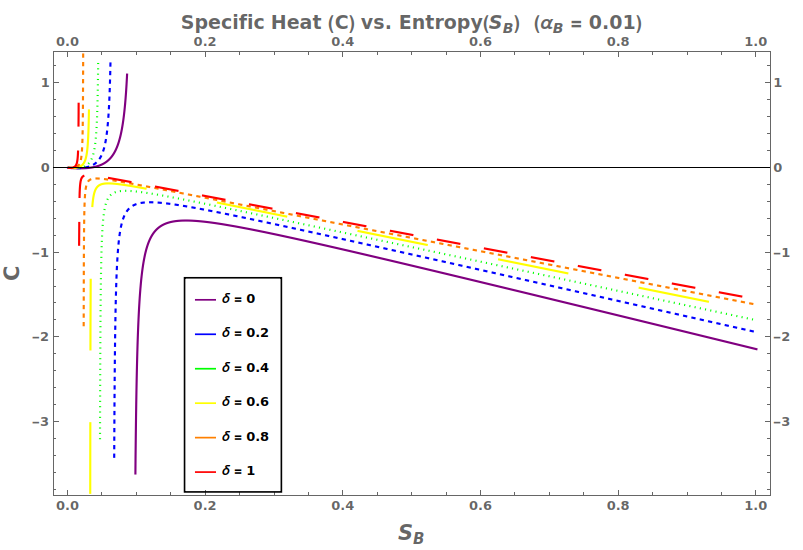}\\
    Figure Caption : fig 2a , 2b and 2c are Heat capacity vs. entropy plots for different values of $\delta$ ($0\leq\delta\leq 1$) in 4DEGB gravity at Gauss Bonnet coupling parameter $\alpha_{GB}=1, ~0.05~ \& ~0.01$ respectively.
  \end{figure}
 Figure 2a to 2c are for $\alpha_{GB} =$ 1, 0.05 and 0.01 respectively. We have shown how specific heat varies with entropy for different values of $\delta$ . For $\alpha_{GB} = 1$, $C$ is a decreasing function, i.e, $C(\delta_1) < C(\delta_2)$ if $\delta_1 > \delta_2$ for the whole range of entropy. As we decrease $\alpha_{GB}$ to 0.05, we find two separate sign branches of C. For $S_B < S_{Bc}(\delta)$, positive specific heat of a stable small black hole is found. Whereas $S_B > S_{Bc}(\delta)$ shows a negative counterpart signifying an unstable one. 
 
The Gibbs free energy can be used to efficiently characterize the phase transition and is expressed as follows:
 \begin{equation}\label{GFE}
F = M - T_H S_B = \frac{c^2}{2 \sqrt{\pi} G_N \left(2 + \delta \right) S_B^{\frac{1}{2 + \delta}}} \left\{ \left(1 + \delta \right) S_B^{\frac{2}{2 + \delta}} + \left(3 + \delta \right) \alpha_{GB} \pi \right\}~~~.
 \end{equation}

\begin{figure}[h!]
    \centering
     ~~~~~~~~~~~~~~~~~~~Fig ~3a ~~~~~~~~~~~~~~~~~~~~~~~~~~~~~~~~~~~~~~~~~~~~~~~~~~~~Fig~3b~~\\
    \includegraphics[width=0.45\linewidth]{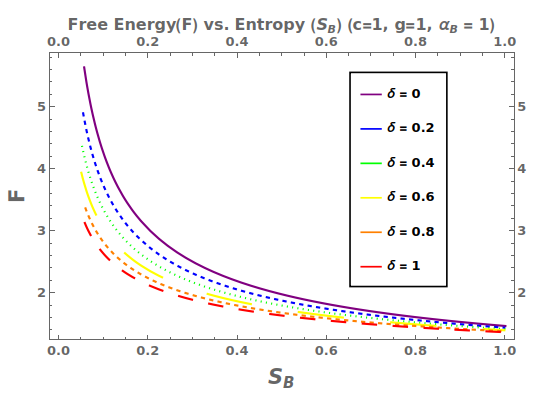}~\includegraphics[width=0.45\linewidth]{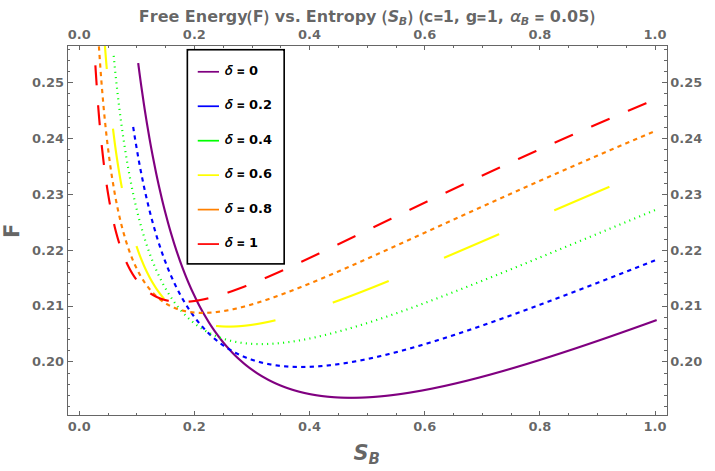}\\  
    ~~~~~~~~~~Fig~3c~~~~~~\\
    \includegraphics[width=0.45\linewidth]{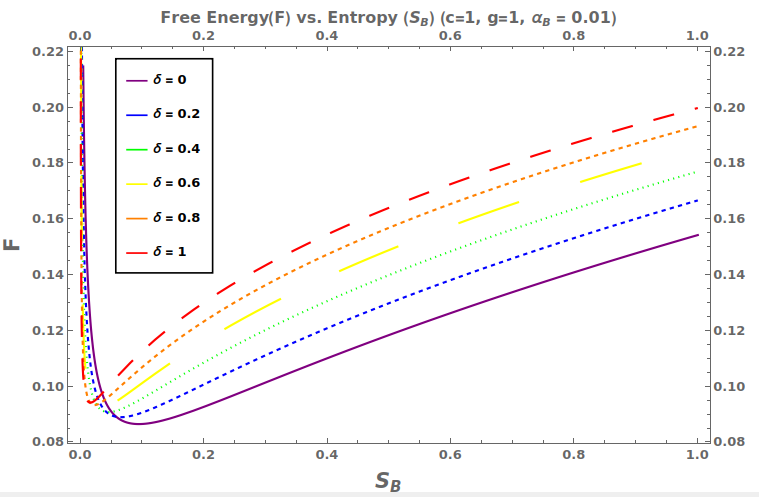}\\
    Figure Caption : fig 3a , 3b and 3c are Free energy vs. entropy plots for different values of $\delta$ ($0\leq\delta\leq 1$) in 4DEGB gravity at Gauss Bonnet coupling parameter $\alpha_{GB}=1, ~0.05~ \& ~0.01$ respectively.
\end{figure}
Free energy  vs.  entropy curves are plotted in 3a to 3c. For $\alpha_{GB} = 1$, in 1a, we observe decreasing free energies $[ F(\delta_1) < F(\delta_2)$  for $\delta_1 >\delta_2 ]$. For $\alpha_{GB} = $ 0.05 and $\alpha_{GB} =$ 0.01, a local maxima is found. It is followed that in $(F,C)$, second  order types changes from $(+,+)$ region to $(+,-)$ region are present. This is the indication of a phase transition. This happens due to existence of thermodynamic quantities to create non equivalent statistical ensembles. To understand this more deeply we will look upon fig 4a to 4c  where free energy is plotted with respect to temperature.

When the effect of $\alpha_{GB}$ is high, in fig 4a, free energy falls as temperature increases. As we reduce the value of $\alpha_{GB}$, in fig 4b and 4c, i.e., we move towards the general relativity, a cuspidal node is found to form. This signifies the second order phase transition. It has been observed that if $\delta_1<\delta_2$, lesser temperature is required for the cuspidal node to formed. 

To classify the nature of phase transition orders, we recall that from (\ref{GFE}), 
\begin{equation}
\frac{dF}{dT_H}=\frac{d}{dT_H}\left(M-T_HS_B\right)=\frac{dM}{dT_H}-\left(S_B+T_H\frac{dS_B}{dT_H}\right)
\end{equation}
\begin{equation}
\implies\frac{d^2F}{dT_H^2}=\frac{d^2M}{dT_H^2}-\left(2\frac{dS_B}{dT_H}+T_H\frac{d^2S_B}{dT_H^2}\right)~~~~.
\end{equation}
The second order derivative of free energy with respect to temperature points how a system will response when temperature is changed. If $\frac{d^2F}{dT_H^2}>0$, the system is passing through a stable state whereas a negative sign indicates instability.

In figure 5a-5c, we plot $\frac{d^2F}{dT_H^2}$ with respect to $S_B$ to follow their continuity properties. For $c=1,~g=1$ and $\alpha_{GB}=1$, this is continuous and hence the transition is of first order. We are confirmed that low $\alpha_{GB}$ ($\geq 0.05$) with Barrow entropy leads to second order phase transitions. More fractal used the entropy, lesser the volume of entropy where the second order phase transition takes place. 

\begin{figure}[h!]
    \centering
     ~~~~~~~~~~~~~~~~~~~Fig ~4a ~~~~~~~~~~~~~~~~~~~~~~~~~~~~~~~~~~~~~~~~~~~~~~~~~~~~Fig~4b~~\\
    \includegraphics[width=0.40\linewidth]{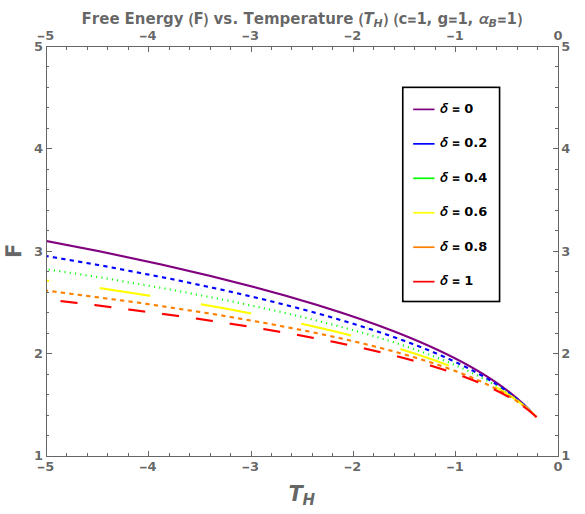}~\includegraphics[width=0.5\linewidth]{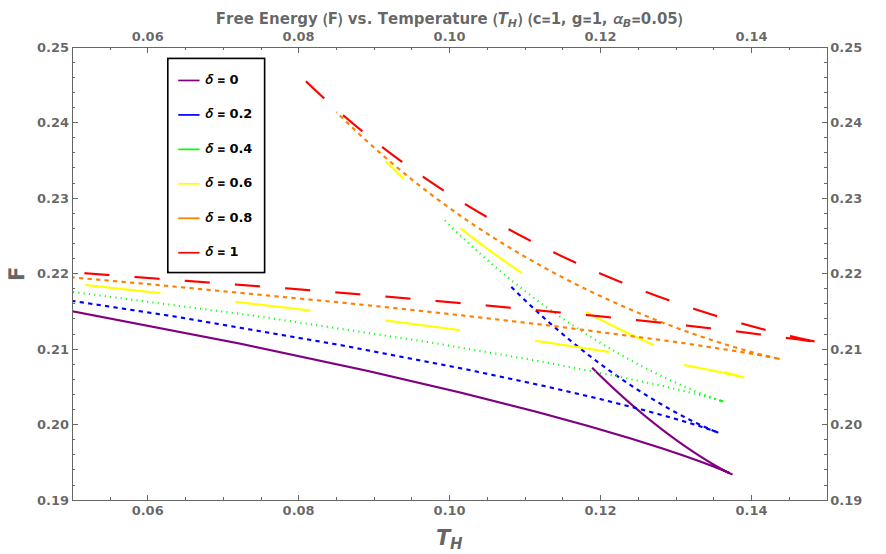}\\  
    ~~~~~~~~~~Fig~4c~~~~~~\\
    \includegraphics[width=0.40\linewidth]{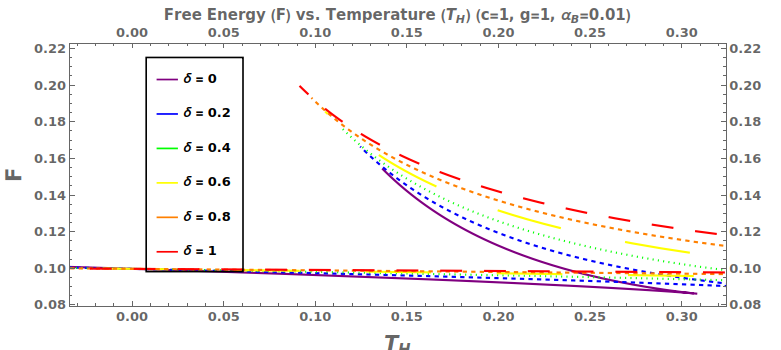}\\
    Figure Caption : fig 4a , 4b and 4c are Free energy vs. Hawking temperature plots for different values of $\delta$ ($0\leq\delta\leq 1$) in 4DEGB gravity at Gauss Bonnet coupling parameter $\alpha_{GB}=1, ~0.05~ \& ~0.01$ respectively.
\end{figure}

\begin{figure}[h!]
    \centering
     ~~~~~~~~~~~~~~~~~~~Fig ~5a ~~~~~~~~~~~~~~~~~~~~~~~~~~~~~~~~~~~~~~~~~~~~~~~~~~~~Fig~5b~~\\
    \includegraphics[width=0.45\linewidth]{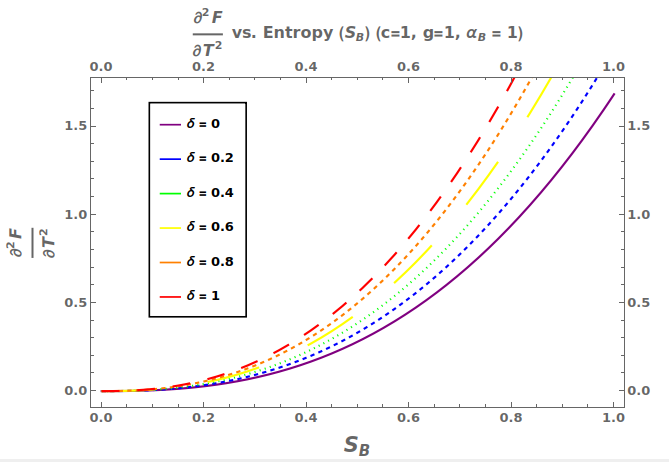}~\includegraphics[width=0.45\linewidth]{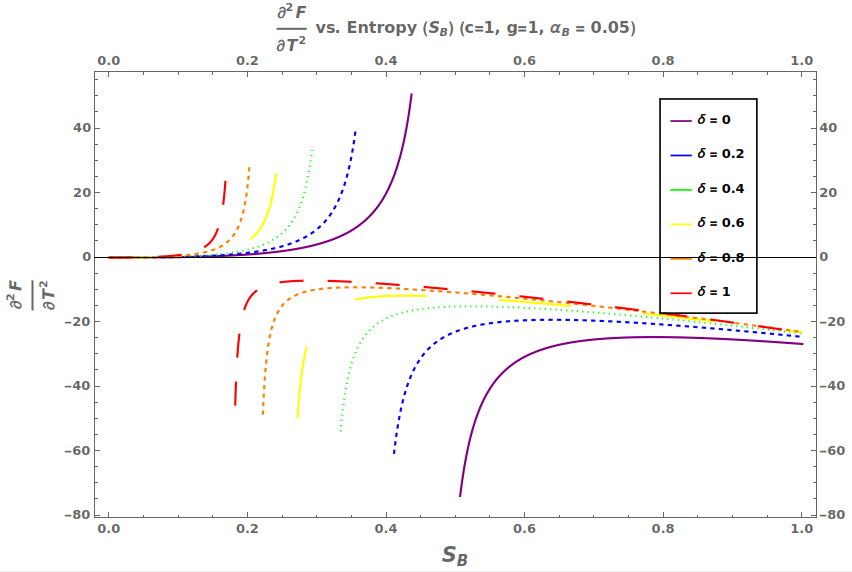}\\  
    ~~~~~~~~~~Fig~5c~~~~~~\\
    \includegraphics[width=0.45\linewidth]{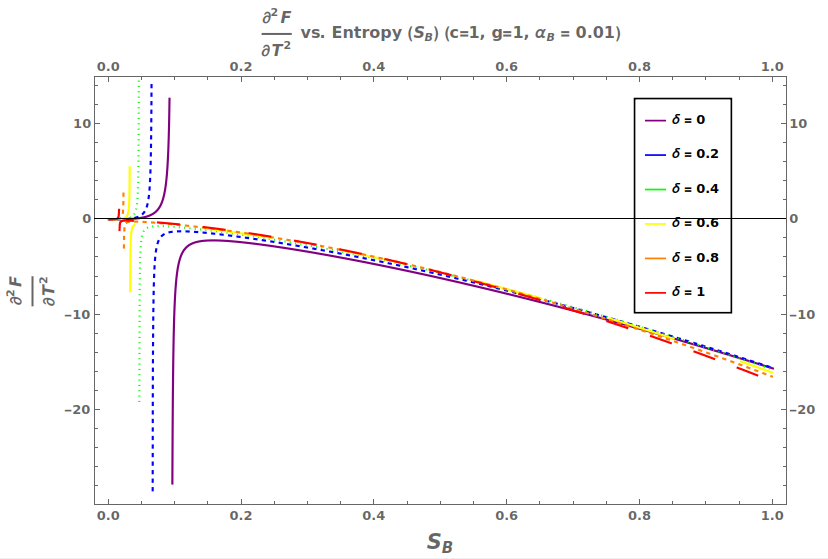}\\
    Figure Caption : fig 5a , 5b and 5c are Free energy vs. $\frac{\partial^2 F}{\partial T^2}$ plots for different values of $\delta$ ($0\leq\delta\leq 1$) in 4DEGB gravity at Gauss Bonnet coupling parameter $\alpha_{GB}=1, ~0.05~ \& ~0.01$ respectively.
\end{figure}

\section{Brief Discussion and Conclusion}
In this article, we have chosen a four dimensional Einstein Gauss Bonnet gravity black hole and applied the fractal modification idea on its entropy. This incorporates one extra parameter to the thermodynamic system, namely $\delta$. It is followed that a four-dimensional Einstein-Gauss-Bonnet gravity black hole possesses only two free parameters , viz. the mass $M$ of the black hole and the Gauss Bonnet coupling parameter $\alpha_{GB}$. We have recalled that to achieve the four dimensional counterpart, we had to either move them a $\alpha_{GB}\rightarrow 4$ limiting process or to reduce the dimension following scalar coupled method. Inclusion of Barrow's crinkled surface increases the information inside the fractalized entropy. This definitely differentiate the whole scenario than a Bekenstein entropy case or a Einstein Gauss Bonnet gravity.

First, we are able to search two horizons(say the event horizon and the Cauchy horizon) for $ M > \frac{c^2}{G_N} \alpha_{GB} $. We find the radii(say $r_h$ and $r_C$ respectively) of these two horizons show a global product $ \frac{c^4}{G_N^2} \alpha_{GB}^2 $ which is free of the mass term. On the similar row, the products of entropies (say $S_{B_h}$ and $S_{B_C}$ respectively) defined on these two horizons turns free of mass. Hence the entropy product law for our system is universal. This kind of double horizon system can not sustain in quantum back reaction background. Existence of double horizon supports a resemblance with classical Reissner Nordstrom or Kerr like black holes.

Next, we have studied the temperature of the related system. This quantity is not a global one as the product over two horizons involves mass explicitly. When $\alpha_{GB}$ parameter is high, i.e., Gauss Bonnet gravity's effect is stronger, temperature is an increasing function. As we reduce the Gauss Bonnet capacity parameter $\alpha_{GB}$, temperature curves are followed to possess two branches with distinct slopes, one steeply increasing and the rest one is slowly decreasing. It is obvious that a local maxima takes place in between. Such a maxima signifies $\frac{\partial T}{\partial S_B}=\frac{\partial^2 M}{\partial S_B^2}=0$ and $\frac{\partial^2 T}{\partial S_B^2}=\frac{\partial^3 M}{\partial S_B^3}<0$. So slowly coupled Gauss-Bonnet gravity demands two distinct phases for a black hole . As we continue to reduce $\alpha_{GB} 's$ value, the  trend of obtaining maxima turns easily notifiable or a second order break in the said branches are followed without a continuous transition. Now keeping $\alpha_{GB}$ fixed, when we have studied the effect of Barrow index on temperature, we find as $\delta$ is increased, the maxima is obtained at low entropy. Hence a fractalized entropy forces the system to transit quickly. We have checked quantum effects on such temperature via generalized uncertainty principle consideration and quantum loop correction. Effect of the first acts very different than the 4D EGB gravity. However, quantum loop correction to the temperature does not ensure any remarkable change.

Next analyzed physical quantity is the specific heat supporting the pattern of temperature variation, specific heat shows there is a smaller unstable black hole at first which passes through a transition to become a larger stable one. As predicted before, this nature takes place for low $\alpha_{GB}$. Free energy and specific heat as a doublet changes sign from $(+, ~-)$ to $(+,~+).$ For smaller  $\alpha_{GB}$, free energy vs. temperature curves are such distinct that the transit even is not smooth. A cusp is formed. hence we conclude that for 4D EGB gravity, black holes are first unstable and then grows to a larger stable counterpart. Fractalization of entropy works as a catalyst for such transitions. The more fractalized the entropy becomes, the less entropy you need for a transition. Fractalized entropy product is found to act as a global quantity. 

\vspace{0.5in}

{\bf Acknowledgment : } RB thanks IUCAA, Pune for granting Visiting Associateship. SP thanks Department of Mathematics, The University of Burdwan for different research facilities.

\bibliographystyle{ieeetr}  
\bibliography{mybib}

\end{document}